\documentstyle{lamuphys}
\include{psfig}
\makeatletter
\let\chapter\hid@chapter
\makeatother

\def\aj{{AJ}}

\def\mnras{{MNRAS}}

\newcommand\CIV{\hbox{C~$\rm IV$}}
\newcommand\HeII{\hbox{He~$\rm II$}}
\def\arcsec{\ifmmode {^{\prime\prime ~}}\else $^{\prime\prime ~}$\fi}
\def\secpoint{\ifmmode \rlap.{^{\prime\prime}}\else
    $\rlap.{^{\prime\prime}}$\fi}
\def\deg{\ifmmode {^{\circ}}\else {$^\circ$}\fi}
\def\eg{{\it e.g.} }
\def\etal{{\it et al.~}}
\def\Lya{{\rm\,Ly-$\alpha$~}}
\def\spose#1{\hbox to 0pt{#1\hss}}
\def\simgt{\mathrel{\spose{\lower 3pt\hbox{$\mathchar"218$}}
     \raise 2.0pt\hbox{$\mathchar"13E$}}}

\begin{document}
\pagenumbering{arabic}
\title{The Highest Redshift Radio Galaxy Known in the Southern Hemisphere}

\author{Carlos De Breuck\inst{1,2}, Wil van Breugel\inst{1}, Huub R\"ottgering\inst{2}, George Miley\inst{2} and Chris Carilli\inst{3}}

\institute{Institute for Geophysics and Planetary Physics, Lawrence Livermore National Laboratory, L$-$413, P.O. Box 808, Livermore, CA 94550, U.S.A.
\and
Leiden Observatory, P.O. Box 9513, 2300 RA Leiden, The Netherlands
\and
National Radio Astronomy Observatory, Socorro, NM ,U.S.A.}
\authorrunning{Carlos De Breuck \etal}

\maketitle

\begin{abstract}
We present the discovery of a $z=4.13$ galaxy TN J1338-1942,
the most distant radio galaxy in the southern hemisphere known to
date. The source was selected from a sample of Ultra Steep Spectrum
(USS; $\alpha < -1.3$; $S \propto \nu^{\alpha}$) radio sources using
the Texas and NVSS catalogs. The discovery spectrum, obtained with
the ESO 3.6m telescope, shows bright extended \Lya emission. The
radio source has a very asymmetric morphology, suggesting a strong
interaction with an inhomogeneous surrounding medium.
\end{abstract}

\section{Southern high redshift radio galaxy searches}
High Redshift Radio Galaxies (HzRGs) may be used to study the
formation and evolution of massive elliptical galaxies (see, \eg van
Breugel, this volume). However, the sample of HzRGs at the highest
redshifts ($z>3$) is extremely small, despite vigorous searches by
several groups. This is especially true in the southern hemisphere: of
the 20 $z>3$ radio galaxies known, only 3 are in the South; below
declination $-40$\deg, only one $z>2$ radio galaxy is known!

To provide samples of HzRGs to study with the soon to be operational
6$-$8m class telescopes in the southern hemisphere (VLT, Gemini$-$South,
Magellan), we have constructed a sample of USS sources from the TEXAS
365 MHz (Douglas \etal 1996) and NVSS 1.4 GHz (Condon \etal 1998)
surveys ($\delta > -35\deg$), and from the MRC 408~MHz (Large \etal
1981) and PMN 4.85~GHz (Griffith \& Wright 1993) surveys ($\delta <
-35\deg$). This USS selection makes our sample $\sim 65\%$ efficient
in selecting $z>2$ radio galaxies (see van Breugel, this volume, and
De Breuck \etal 1998).

\section{The first southern $z>4$ radio galaxy}
The highest redshift USS object we have found thus far is the radio
galaxy TN~J1338-1942. The source has an integrated spectral index
$\alpha = -1.33$, and a straight power-law spectrum between 365~MHz,
1.4~GHz and 4.8~GHz. High resolution VLA imaging at 4.8~GHz and
8.3~GHz shows that the source has a core with $\alpha_{8.3}^{4.8} =
-0.6$, a northwestern lobe with $\alpha_{8.3}^{4.8} = -1.6$ at
1\arcsec from the core, and a southeastern lobe with
$\alpha_{8.3}^{4.8} = -2.5$ at 3\secpoint6 from the core (Fig. 1). The
coincidence of extended \Lya~ emission with the NW lobe suggests that
this asymmetry may be due to an inhomogeneous ambient medium.

The $R-$band identification and spectroscopic observations were
obtained with the EFOSC1 imaging spectrograph on the ESO 3.6m
telescope. We obtained two spectra, one covering 3725\AA~ to 6940\AA\
(not shown here), and another covering 6000\AA~ to 9200\AA\
(Fig. 2). The 2\arcsec slit used to obtain the blue spectrum was
offset 2\arcsec from the radio core, but nevertheless showed bright
\Lya, proving the large extent of the \Lya~ emission.

\vspace{0.2cm}
\hspace{-1cm}
\begin{minipage}{6cm}
\psfig{file=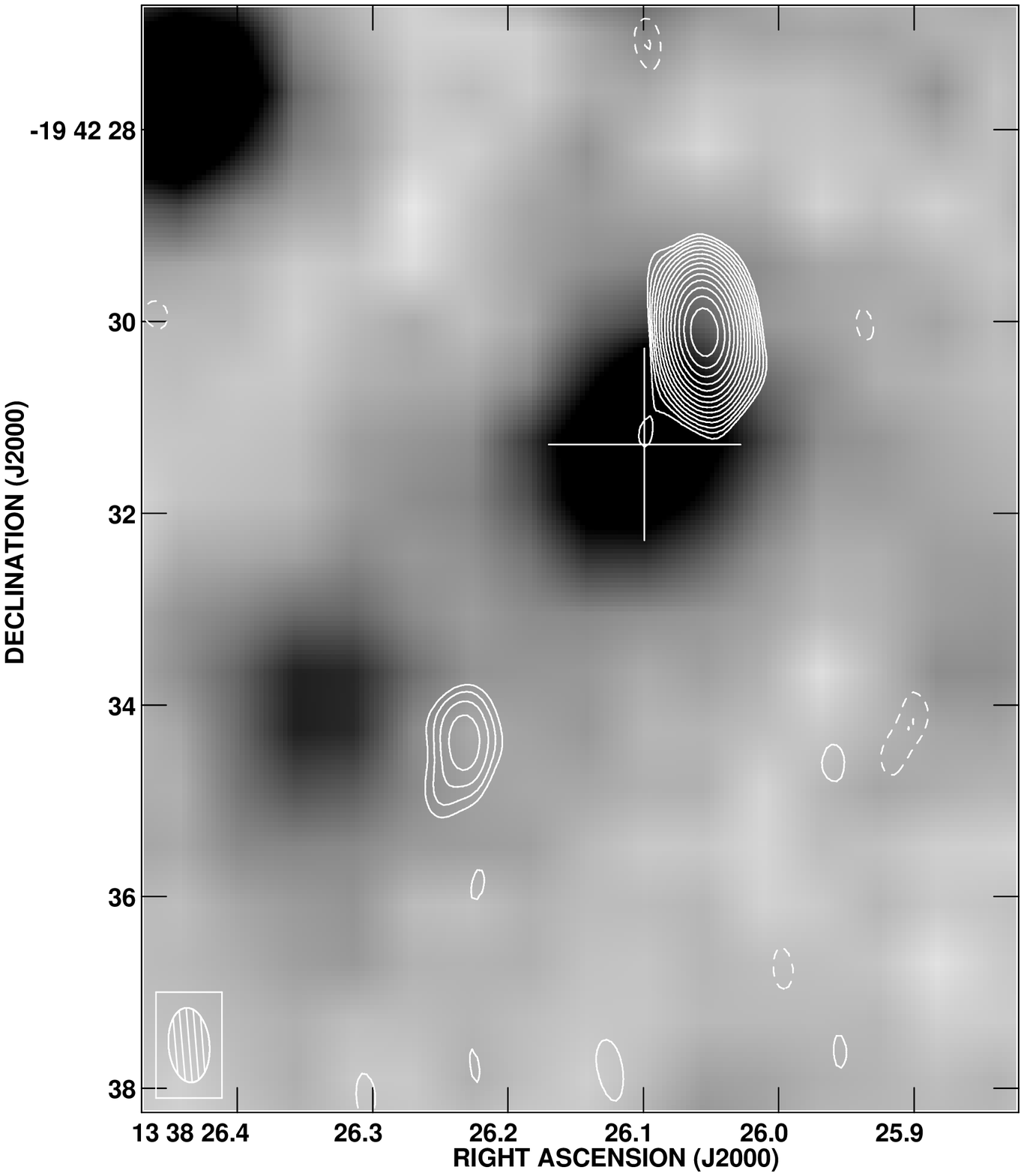,height=6.5cm}
{\bf Figure 1.}
\small{{\it Greyscales:} $R-$band image ($t_{int}$~=~10~min, 1\secpoint3 seeing), dominated by \Lya~ emission in the passband. {\it Contours:} VLA 4.71~GHz map showing the asymmetric lobes. The flat-spectrum core is indicated by a cross.}
\end{minipage}
\hspace{0.4cm}
\begin{minipage}{6cm}
\hspace{-0.9cm}
\psfig{file=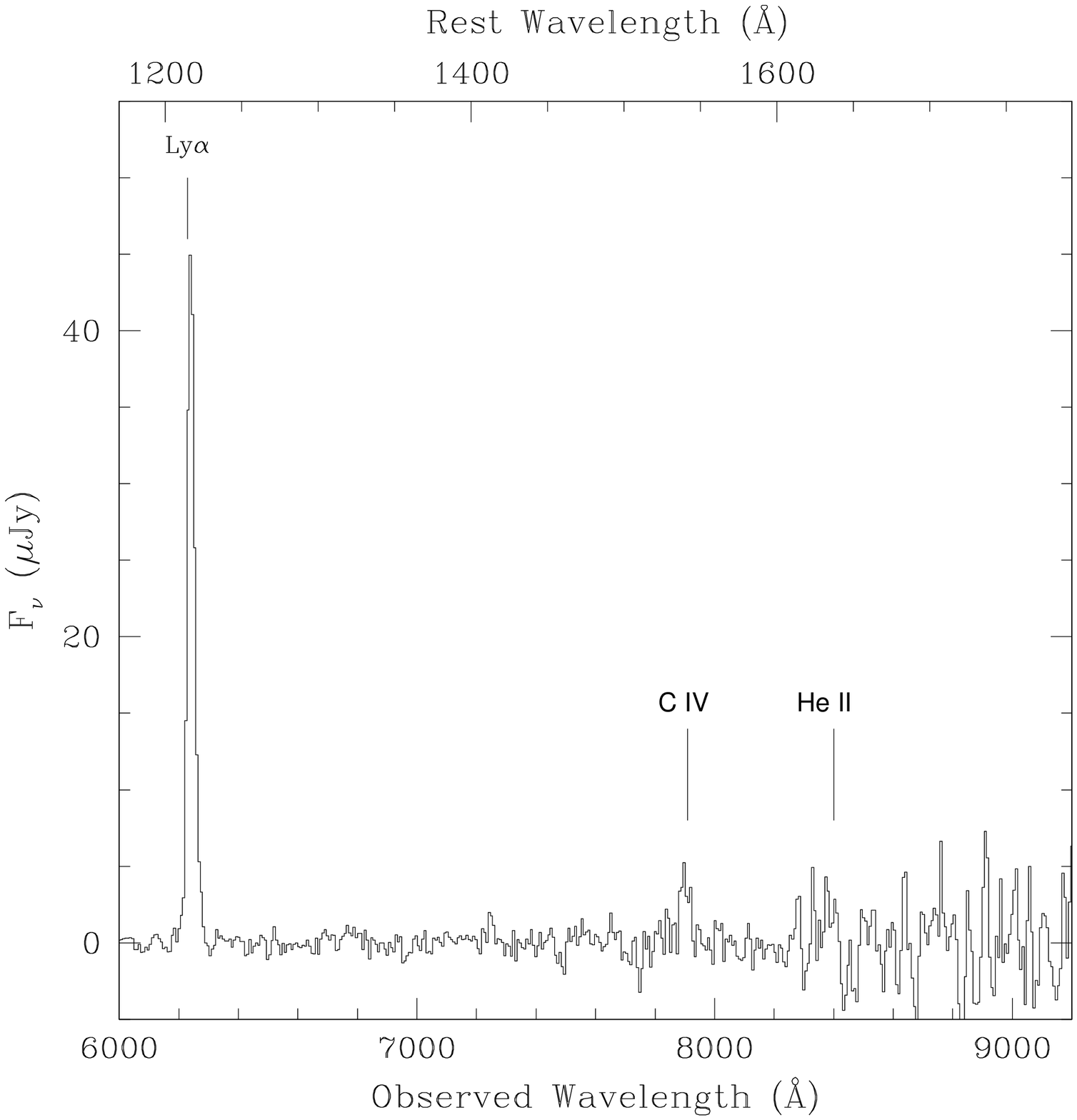,height=6.0cm,width=6.8cm}

{\bf Figure~2.}
\small{ESO~3.6m~spectrum~of TN~J1338$-$1942~($t_{int}$~=~105~min).~~~The bright ($1 \times 10^{-15}$~erg~s$^{-1}$cm$^{-2}$) \Lya is extended (6\arcsec; $\sim$ 50~kpc) and has spatially extended absorption on the blue side, as seen in many $z > 3$ radio galaxies (\eg\ Dey 1998). Weak \CIV~ and \HeII~ lines confirm the redshift $z=4.13$.}
\end{minipage}

\vspace{0.5cm}
\noindent {\bf Acknowledgments} The research at IGPP/LLNL is performed
under the auspices of the US Department of Energy under contract
W--7405--ENG--48.

%
%

\end{document}